\documentclass[pre,twocolumn,a4paper,superscriptaddress]{revtex4}
\usepackage{graphicx,amssymb}
\begin{document}
%My commands
\newcommand{\be}{\begin{equation}}
\newcommand{\ee}{\end{equation}}
\newcommand{\bq}{\begin{eqnarray}}
\newcommand{\eq}{\end{eqnarray}}
\newcommand{\bsq}{\begin{subequations}}
\newcommand{\esq}{\end{subequations}}
\newcommand{\bc}{\begin{center}}
\newcommand{\ec}{\end{center}}
\newcommand {\R}{{\mathcal R}}
\newcommand{\al}{\alpha}
\newcommand\lsim{\mathrel{\rlap{\lower4pt\hbox{\hskip1pt$\sim$}}
    \raise1pt\hbox{$<$}}}
\newcommand\gsim{\mathrel{\rlap{\lower4pt\hbox{\hskip1pt$\sim$}}
    \raise1pt\hbox{$>$}}}

\title{Von-Neumann's and related scaling laws in Rock-Paper-Scissors type models}
\author{P.P. Avelino}
%\email[Electronic address: ]{ppavelin@fc.up.pt}
\affiliation{Centro de Astrof\'{\i}sica da Universidade do Porto, Rua das Estrelas, 4150-762 Porto, Portugal}
\affiliation{Departamento de F\'{\i}sica e Astronomia, Faculdade de Ci\^encias, Universidade do Porto, Rua do Campo Alegre 687, 4169-007 Porto, Portugal}
\author{D. Bazeia}
%\email[Electronic address: ]{bazeia@fisica.ufpb.br}
\affiliation{Departamento de F\'{\i}sica, Universidade Federal da Para\'{\i}ba 58051-970 Jo\~ao Pessoa, Para\'{\i}ba, Brazil}
\author{L. Losano}
%\email[Electronic address: ]{losano@fisica.ufpb.br}
\affiliation{Departamento de F\'{\i}sica e Astronomia, Faculdade de Ci\^encias, Universidade do Porto, Rua do Campo Alegre 687, 4169-007 Porto, Portugal}
\affiliation{Departamento de F\'{\i}sica, Universidade Federal da Para\'{\i}ba 58051-970 Jo\~ao Pessoa, Para\'{\i}ba, Brazil}
\affiliation{Centro de F\'{\i}sica do Porto, Rua do Campo Alegre 687, 4169-007 Porto, Portugal}

\author{J. Menezes}  
%\email[Electronic address: ]{jmenezes@ect.ufrn.br} 
\affiliation{Departamento de F\'{\i}sica e Astronomia, Faculdade de Ci\^encias, Universidade do Porto, Rua do Campo Alegre 687, 4169-007 Porto, Portugal}
\affiliation{Centro de F\'{\i}sica do Porto, Rua do Campo Alegre 687, 4169-007 Porto, Portugal} 
\affiliation{Escola de Ci\^encias e Tecnologia, Universidade Federal do Rio Grande do Norte\\
Caixa Postal 1524, 59072-970, Natal, RN, Brazil}


\date{\today}
\begin{abstract}
We introduce a family of Rock-Paper-Scissors type models with $Z_N$ symmetry ($N$ is the number of species) and we show that it has a very rich structure with many completely different phases. We study realizations which lead to the formation of domains, where individuals of one or more species coexist, separated by interfaces whose (average) dynamics is curvature driven. This type of behavior, which might be relevant for the development of biological complexity, leads to an interface network evolution and pattern formation similar to the ones of several other nonlinear systems in condensed matter and cosmology.
\end{abstract}
%\pacs{}
%\keywords{Cosmology; Dark energy}
\maketitle

The mechanisms leading to the enormous biodiversity observed in nature are still not fully understood. Rock-Paper-Scissors (RPS) type models incorporate some of the crucial ingredients associated with the dynamics of a network of competing species and they have been used as a powerful tool in the study of complex biological systems. In its simplest form, the RPS game describes the evolution of 3 species which cyclically dominate each other \cite{Kerr2002,Reichenbach2007,Shi2010} (see also \cite{Volterra,doi:10.1021/ja01453a010} for the pioneer work by Lotka and Volterra). The basic interactions are Motion, Reproduction and Predation but generalizations, incorporating more than 3 species and/or new interactions between them, have been proposed in the literature \cite{PhysRevE.76.051921,Peltomaki2008,Szabo2008,Wang2010,Wang2011,Hawick2011,Hawick_2011,1101.0018,PhysRevE.85.051903}. 
Particularly, in \cite{PhysRevE.76.051921} segregation processes and phase transitions in predator-prey models with an even number of species have been investigated. 

RPS type models naturally lead to the formation of complex spatial patterns observed in some biological systems \cite{Kerr2002}. Complex spatial structures also arise in many other systems. For example, interfaces in ideal soap froths and grain growth have a velocity $v$ proportional to the mean curvature $\kappa$ at each point, which is at the core of the Von Neumann's law \cite{vonNeumann1952} for the evolution of the area of individual domains. The evolution with time $t$ of the characteristic scale $L$ of such interface networks obeys the scaling law $L \propto t^{1/2}$, leading to the formation of cellular patterns of increasing size \cite{Glazier-1989-PhdThesis,Stavans1989,Glazier1992,Flyvbjerg1993,Monnereau1998,Weaire2000,Kim2006,PhysRevLett.98.145701}. Despite the different dynamical scaling laws, the evolution of interface networks in a cosmological context may also generate similar spatial patterns to the ones observed in soap froths \cite{Avelino2008,Avelino2010}. 

Consider a family of models where individuals of various species are distributed on a square  lattice of size ${\mathcal N}$ at some initial time. The different species are labelled by the number $i$ (or $j$) with $i,j=1,...,N$, and we make the cyclic identification $i=i+k\,N$ where $k$ is an integer. The number of individuals of the species $i$ will be denoted by $I_i$. In addition to individuals, there are also empty spaces which shall be denoted by $E$. In this paper, except if stated otherwise, we shall assume that the initial distribution is random and that the number densities of the various species $n_i=I_i/{\mathcal N}$ are all identical at the initial time. The number density of empty spaces is given by $n_E = I_E/{\mathcal N}$ and its initial value is equal to $0.1$ in all the network simulations described in this paper. At each time step a random individual (active) interacts with one of its four nearest neighbors (passive).  The unit of time $\Delta t=1$ is defined as the time necessary for  ${\mathcal N}$ interactions to occur (one generation time). The possible interactions are classified as Motion 
$$
iE \to Ei\,,  \quad {\rm or} \quad ij \to ji\,,
$$ 
Reproduction 
$$
iE \to ii\,,
$$ 
or Predation 
$$
i(i-\alpha) \to iE\,, {\rm or} \, i(i+\alpha) \to iE\,,
$$
where $\alpha=1,...,\alpha_{max}$ with $\alpha_{max}=N/2$ if $N$ is even or $\alpha_{max}=(N-1)/2$ if $N$ is odd. We shall denote the corresponding probabilities by $m_i$ (Motion), $r_i$ (Reproduction), $p_{Li\alpha}$ (left-handed Predation) and $p_{Ri\alpha}$ (right-handed Predation). This family of models has a $Z_N$ symmetry if $m_i=m$, $r_i=r$, $p_{Li\alpha}=p_{L\alpha}$ and $p_{R i\alpha}=p_{R\alpha}$ for all $i$. Figure 1 represents the different predation interactions in a model with $5$ species having $Z_5$ symmetry. Throughout this paper, we shall assume that ${\mathcal N}=600^2$ and that the $Z_N$ symmetry is preserved with the following interacting probabilities: $m=0.5$, $r=0.25$ (or zero, if the passive is not $E$), $p=0.25$ (or zero, if the passive is $E$ or if the passive is not a prey of the active). The $Z_N$ symmetry is, in general, not associated with curvature driven dynamics. For example, the standard RPS model has a $Z_3$ symmetry but the dynamics of the spatial patterns is not curvature driven in this model. In fact, we shall show that the dynamics is curvature driven only in realizations which result in the formation of domains, where individuals of one or more species coexist, separated by interfaces whose dynamics is controlled by interactions of equal strength between competing species.

Let us start by introducing a simple model with $N=2$ having $p_{L1}=p_{R1} \neq 0$ (model I). This model has 2 competing species which tend to distribute themselves into separate domains bounded by interfaces where most of the action occurs. Predation happens mainly at the interfaces whose thickness is directly related to the mobility. This is clearly shown on Fig. 2 where a snapshot of the network evolution of model I is presented. On the left panel each species is represented by a different color while on the right panel the black dots represent the distribution of empty spaces. Fig. 2 clearly shows that the empty spaces, which are a result of Predation between individuals of the competing species, are located at the domain borders. The average thickness $\epsilon$ of the interfaces does not change with time.

%%%%%%%%%%%%%%%%%%%%%%%%%%%%%%%%%%%%%%%%%%%%%%
\begin{figure}
\includegraphics*[width=6.2cm, height=6.2cm]{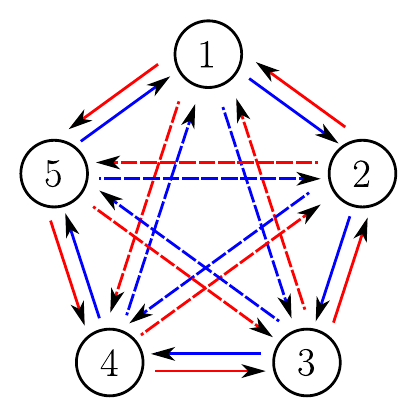}
\caption{(Color online) Predation interactions in a model with $5$ species having $Z_5$ symmetry.}
\end{figure}
%%%%%%%%%%%%%%%%%%%%%%%%%%%%%%%%%%%%%%%%%%%%%%

For an interface with curvature radius $\rho$ and thickness $\epsilon \ll \rho$, the average number of attacks per unit time from individuals outside the border is proportional to the outer interface length (proportional to $\rho+\epsilon/2$) while the average number of attacks from individuals inside the border is proportional to the inner  interface length  (proportional to $\rho-\epsilon/2$). The difference between the average number of attacks per unit of time from outside and inside the border is proportional to $\epsilon$. On the other hand, the number of attacks necessary to modify the domain radius by $\Delta \rho \ll \rho$ is proportional to the interface length ($\propto \rho$). This implies that the value of the velocity of the interface is on average proportional to its curvature ($v=C\kappa=C \rho^{-1}$, where $C$ is a positive constant), which is typical of non-relativistic interfaces in condensed matter \cite{Avelino2010}.

%%%%%%%%%%%%%%%%%%%%%%%%%%%%%%%%%%%%%%%%%%%%%%
\begin{figure}
\includegraphics*[width=4.2cm, height=4.2cm]{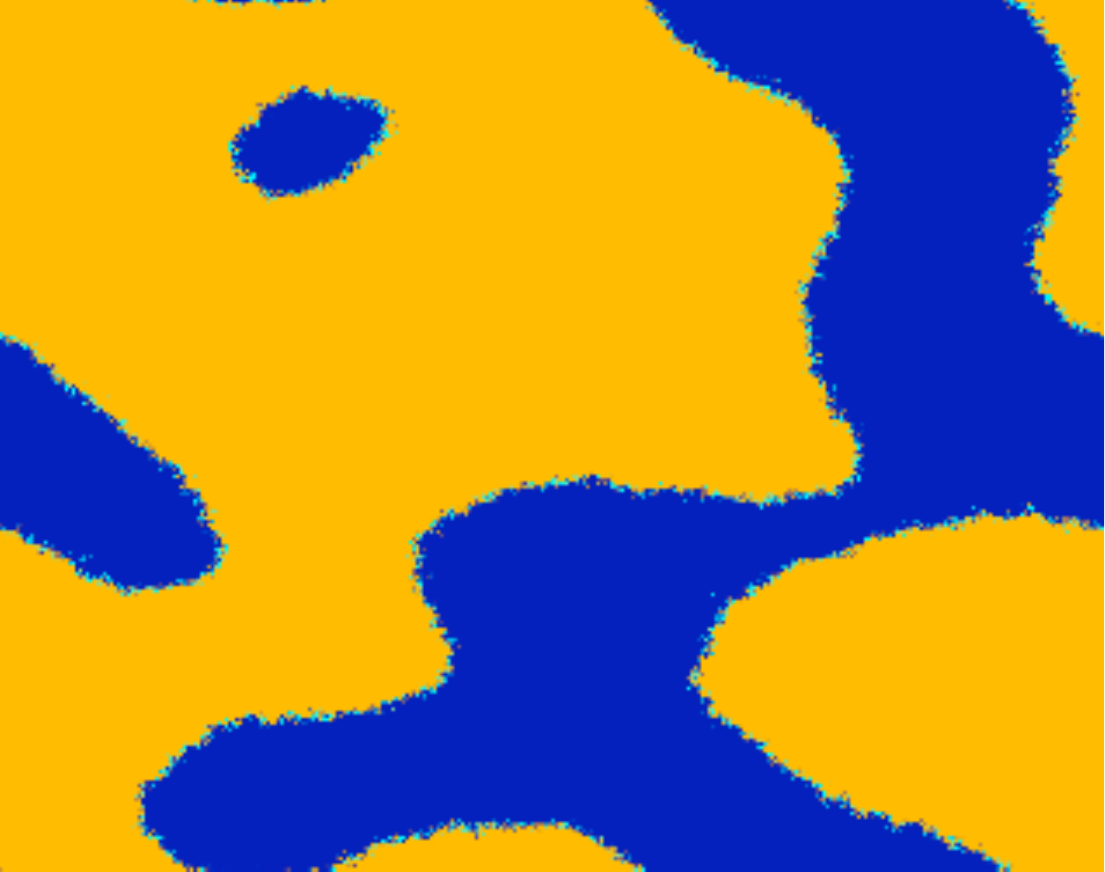}
\includegraphics*[width=4.2cm, height=4.2cm]{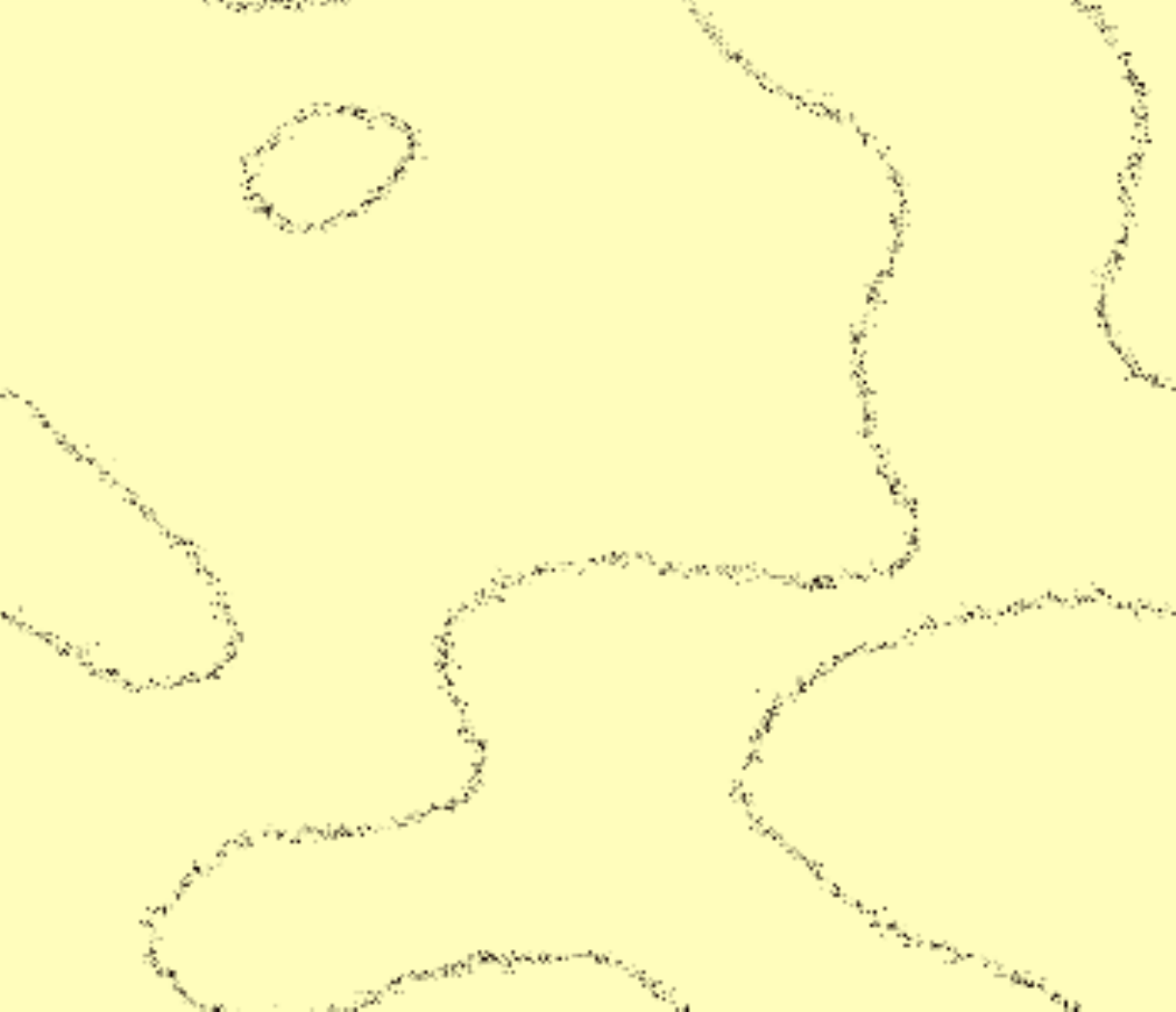}
\caption{\label{z2} (Color online) Snapshot of the evolution of model I. The 2 species are represented using different colors on the left panel. On the right panel the black dots represent the distribution of empty spaces.}
\end{figure}
%%%%%%%%%%%%%%%%%%%%%%%%%%%%%%%%%%%%%%%%%%%%%%

The average evolution of the area of a compact simply connected domain with no vertices is then given by ${\dot a}=\oint v dl=-C \oint \kappa dl=-2\pi C$, where $dl$ is the infinitesimal interface arc length and a dot denotes a derivative with respect to $t$. Note that the domain area can be calculated at any given time by counting the number of individuals inside the domain. If the domain is compact, but not necessarily simply connected, then
\be
{\dot a}_{[g]} = 2 \pi C (g-1)\,,
\label{dandt1}
\ee
where ${a}_{[g]}$ represents the area of a compact domain with genus $g$ and no vertices. Eq. (\ref{dandt1}) implies that the area decreases (increases) with time depending on whether $g=0$ ($g>0$). The genus dependency accounts for the contribution of the decrease of the area of each hole to the growth of the area of the domain. If $g=0$ then, on average, the evolution of the area with time is given by \be
a(t)=a(0) \left(1-t/t_c\right)\,,
\label{area}
\ee
where $t_c$ is the time of collapse. The average time evolution of the area $a(t)$ of an initial circular domain with $g=0$ is illustrated in Fig. 3. The domain area $a(t)$ was calculated by counting the number of individuals of the species inside the domain. We verified that a nearly identical result is obtained using the relation $a \propto I_E^2$. The solid red line represents the average result in an ensemble of 43 simulations using model I and the dashed blue line shows the theoretical evolution given by Eq. (\ref{area}), with $t_c$ calculated as the median collapse time. Fig. 3 shows that the agreement between the analytical and numerical results is very good.

In the case of an interface network without junctions the evolution of the total number of domains, $N_D$, with time can be obtained using Eq. (\ref{dandt1}). If the fraction $f_{[0]}$ of the total number of domains with genus $g=0$ is a constant (${\dot f}_{[0]}=0$) then \cite{Glazier-1989-PhdThesis}
\be
\label{dntdt}
\frac{{\dot N}_D}{N_D}= \frac{{\dot a}_{[0]}}{{\bar a}_{[0]}} f_{[0]}\,.
\ee
Here ${\bar a}_{[g]}=h_{[g]} {\bar a}$ is the average area of a compact domain with genus $g$, ${\bar a}=A/N_D$ is the average domain area, $A=\sum_{g=0}^\infty A_{[g]}$ is the area of the entire system and $A_{[g]}$ is the total area occupied by domains with genus $g$. If both $f_{[0]}$ and $h_{[0]}$ are time independent, then ${\dot N}_D \propto  - N_D^2$. The characteristic length of the network defined by $L \equiv (A/N_D)^{1/2}$ evolves as
\be
L \propto N_D^{-1/2} \propto t^{1/2}\,.
\label{lev}
\ee

%%%%%%%%%%%%%%%%%%%%%%%%%%%%%%%%%%%%%%%%%%%%%%%%%%%%%%%%%%%%%
\begin{figure}
\begin{center}
\includegraphics*[width=7.4cm]{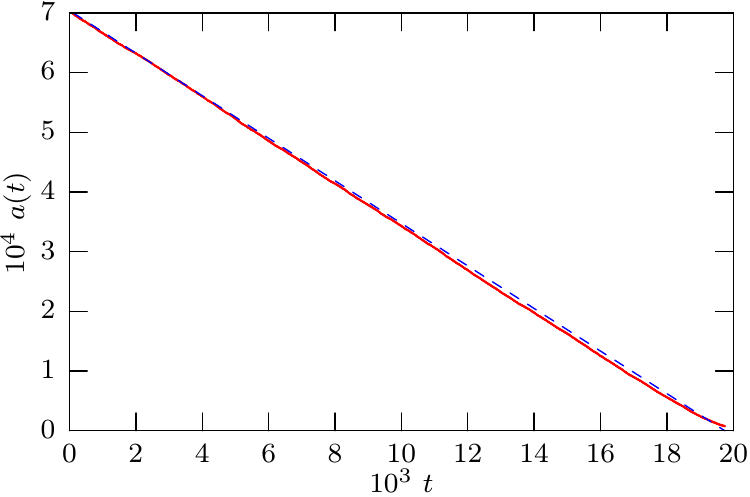}
\end{center}
\caption{(Color online) The time evolution of the area of a circular domain. The solid red line represents the average result in an ensemble of 43 simulations using model I and the dashed blue line is the theoretical prediction.}
\label{circular}
\end{figure}
%%%%%%%%%%%%%%%%%%%%%%%%%%%%%%%%%%%%%%%%%%%%%%%%%%%%%%%%%%%%%%%%%

Fig. 4 shows the evolution of the number of empty spaces $I_E$ with time $t$ for model I. The cyan dots (light grey in black and white) represent the results of 20 interface network simulations while the solid line represents the average. The scaling exponent $\lambda$ defined by $I_E \propto t^{\lambda}$ is $\lambda = -0.450 \pm 0.027$ at one-sigma. The number of domains $N_D$ is equal to the ratio between the total area $A$ and the average domain area $L^2$. It is also proportional to the ratio between the total interface length $L_T$ and the average domain perimeter which is proportional to $L$. Hence, $A/L^2 \propto L_T/L$ implying that $L \propto A/L_T$. On the other hand, the average thickness $\epsilon$ of the interfaces does not change with time and consequently the total interface length $L_T$ is proportional to $I_E$. Hence, $L$ is inversely proportional to the number of empty spaces ($L \propto I_E^{-1}$) and consequently one would expect an average scaling solution with $I_E \propto t^{\lambda}$, with $\lambda=-0.5$. The fact that the numerical value of $\lambda$ is very close to the theoretical one demonstrates that the interface network evolution is already attaining the expected scaling regime. The increase with time of the dispersion between the values of $I_E$ for the various simulations is associated with the growth of the characteristic length scale of the network (see \cite{Avelino2005} for a detailed analytical discussion).

If $v = C \kappa$ then the evolution of the area of a single domain with genus $g=0$ and $\ell$ vertices is given by 
\be
{\dot a}_\ell = - C \oint \kappa dl =- C \left(2\pi - \sum_{\beta=1}^\ell \theta_\beta\right)\,,
\ee
where $\theta_\beta$ represent each of the $\ell$ discontinuous angle changes at the vertices. If the interface network has only Y-type junctions which meet at an angle of $2 \pi/3$, then one obtains the Von Neumann's law \cite{vonNeumann1952}
\be
{\dot a}_\ell=- C \left[2\pi - \ell\left(\pi-\frac{2\pi}{3}\right)\right]=C\frac{\pi}{3} (\ell-6)\,,
\label{dandt}
\ee
implying that the area domains with $\ell<6$ ($\ell>6$) decreases (increases) proportionally to time. The evolution of interface networks in RPS type models is not deterministic and consequently the Von Neumann's law can only be valid on average. Fig. 5 shows the evolution of a Y-type junction in model IV where $p_{L\alpha}=p_{R\alpha}$ for all $\alpha$ (this model will be described later in more detail), starting from the initial configuration shown in the left panel. The right panel represents the most frequent species at each lattice point from $t=1 \times 10^5$ to $t=2.5  \times 10^5$, showing that the average angles at the vertex all tend to an average value of $2\pi/3$. 

%%%%%%%%%%%%%%%%%%%%%%%%%%%%%%%%%%%%%%%%%%%%%%%%%%%%%%%%%%%%%
\begin{figure}
\begin{center}
\includegraphics*[width=7.4cm]{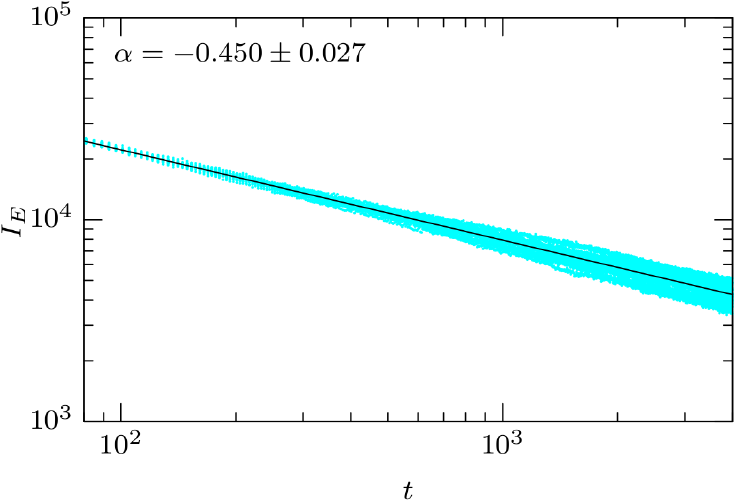}
\end{center}
\caption{(Color online) Evolution of the number of empty spaces $I_E$ with time $t$ for model I. The cyan dots (light grey in black and white) represent the results of 20 interface network simulations while the solid line represents the average. The scaling exponent $\lambda$ defined by $I_E \propto t^{\lambda}$ is estimated as $\lambda = -0.450 \pm 0.027$.}
\label{evie}
\end{figure}
%%%%%%%%%%%%%%%%%%%%%%%%%%%%%%%%%%%%%%%%%%%%%%%%%%%%%%%%%%%%%%%%%

The evolution of the total number of domains $N_D$ with time, in the case of an interface network with Y-type junctions, can be obtained using Eq. (\ref{dandt}). If the fraction $f_\ell$ of the total number of domains with $\ell$ edges is a constant (${\dot f}_\ell=0$) then
\be
\label{dntdt1}
\frac{{\dot N}_D}{N_D}= \sum_{\ell < 6} \left(\frac{{\dot a}_\ell}{{\bar a}_\ell} w_\ell\right)\,.
\ee
Here ${\bar a}_\ell=h_\ell {\bar a}$ is the average area of a domain with $\ell$ edges, ${\bar a}=A/N_D$ is the average domain area, $A=\sum_{\ell=1}^\infty A_\ell$ is the area of the entire system, $A_\ell$ is the total area occupied by domains with $\ell$ edges and the function $w_\ell$ accounts for the fact that the collapse of an individual domain might lead to the merger of some of the surrounding domains ($w_\ell=1$ if all domains are of different types and $w_\ell > 1$ otherwise). If the interface network is in a scaling regime with time-independent $f_\ell$, $h_\ell$ and $w_\ell$, then again ${\dot N}_D \propto  - N_D^2$.  Consequently, the scaling law given in Eq. (\ref{lev}) also applies to interface networks with Y-type junctions.

%%%%%%%%%%%%%%%%%%%%%%%%%%%%%%%%%%%
\begin{figure}
\includegraphics*[width=4.2cm, height=4.2cm]{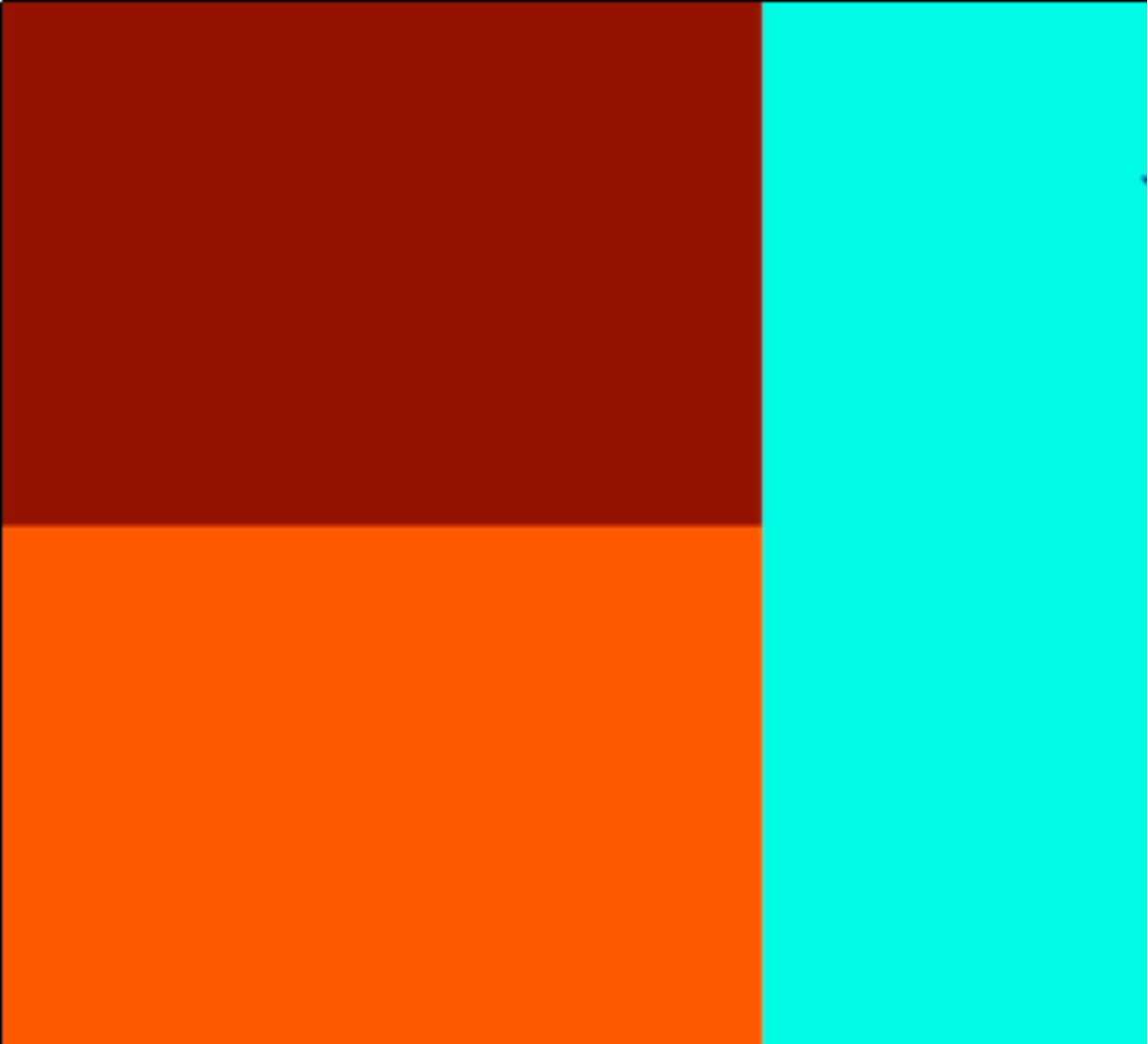}
\includegraphics*[width=4.2cm, height=4.2cm]{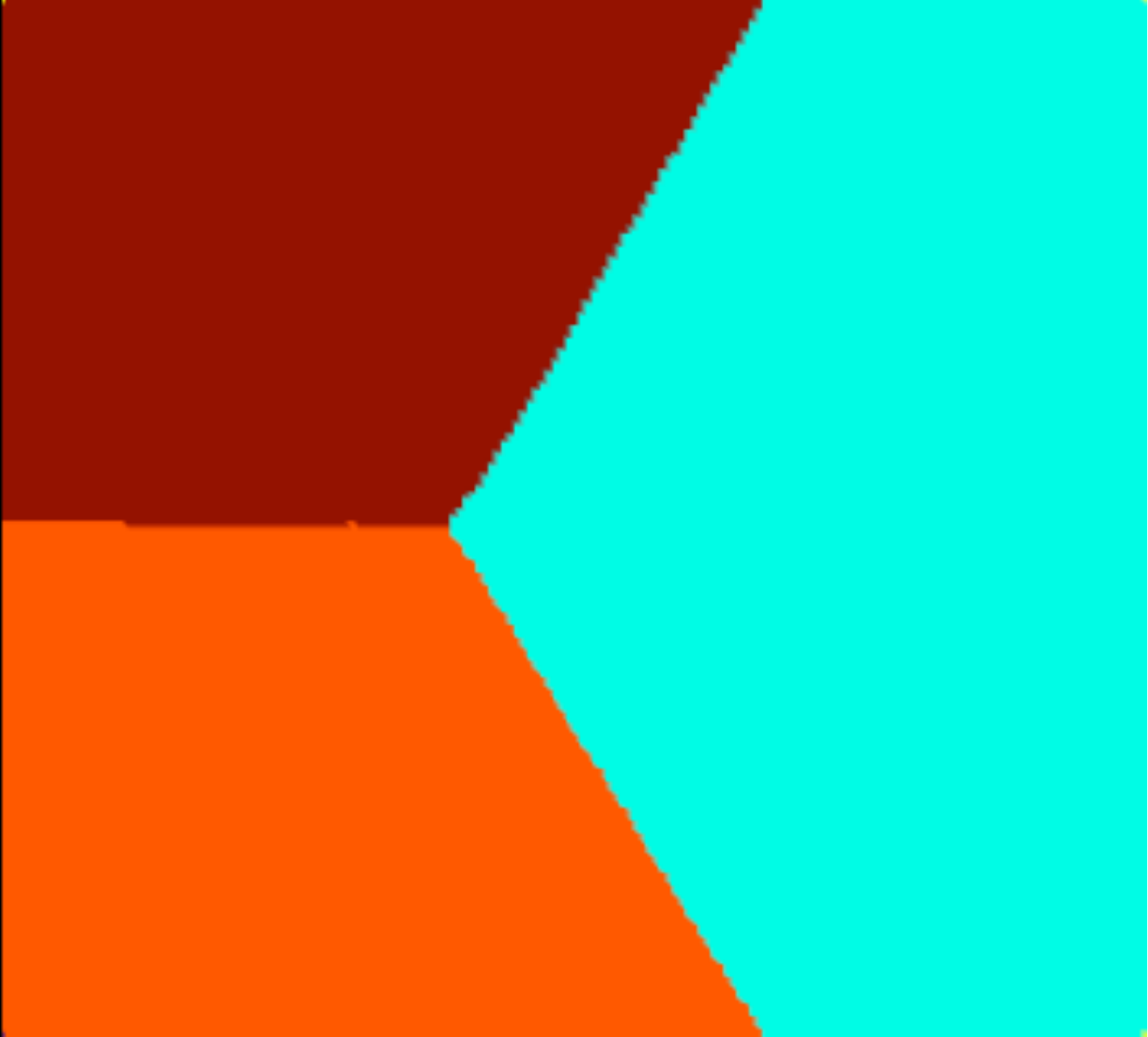}
\caption{(Color online) Evolution of a Y-type junction in model IV where $p_{L\alpha}=p_{R\alpha}$ for all $\alpha$. The left panel shows the initial configuration with different species represented by different colours. The right panel represents the most frequent species at each point from $t=1 \times 10^5$ to $t=2.5\times 10^5$ (note that the average angles at the vertex are all approximately equal to $2 \pi/3$). }
\end{figure}
%%%%%%%%%%%%%%%%%%%%%%%%%%%%%%%%%%%

%%%%%%%%%%%%%%%%%%%%%%%%%%%%%%%%%%%%%%%%%%%%%%
\begin{figure}
\includegraphics*[width=4.2cm, height=4.2cm]{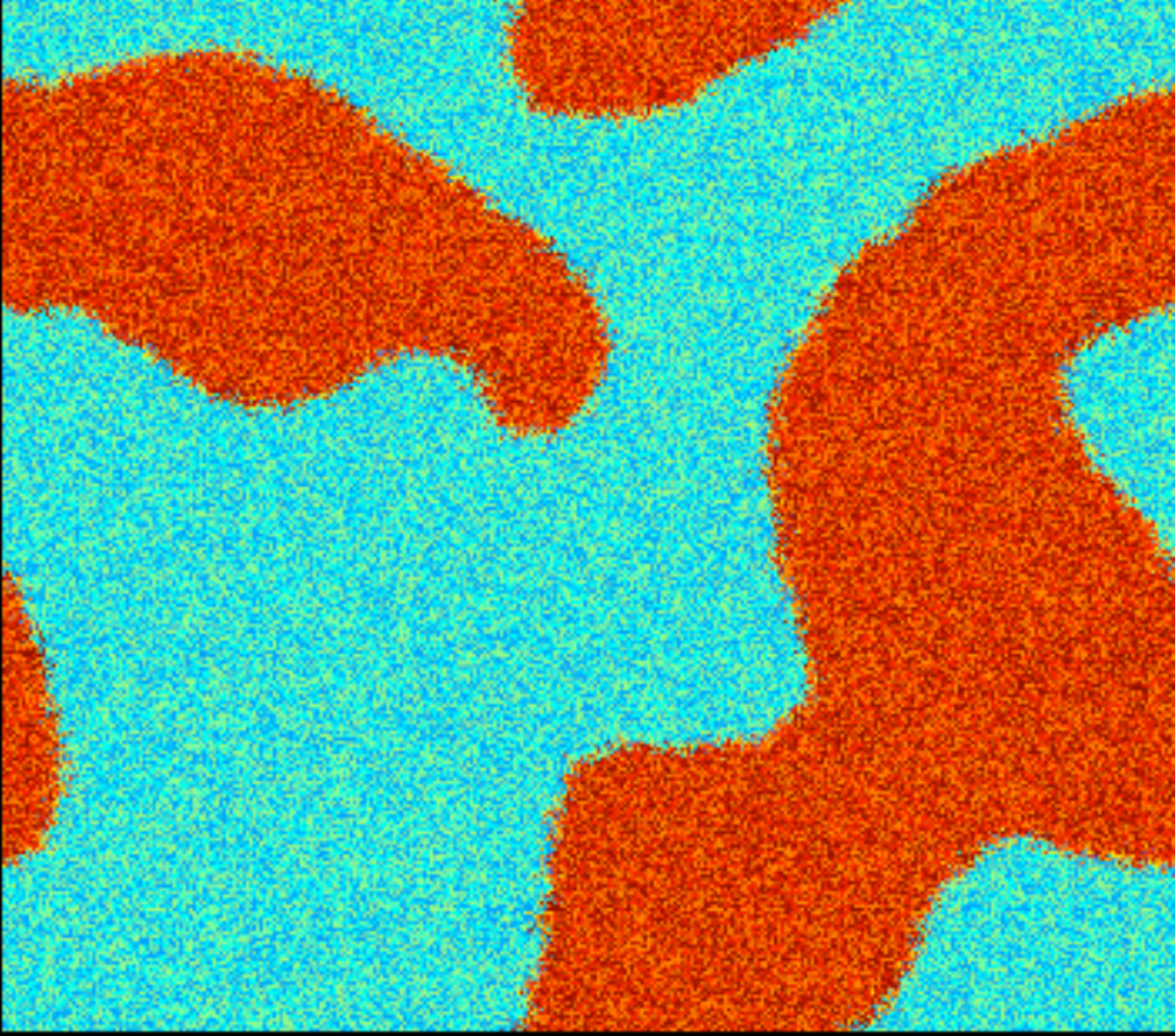}
\includegraphics*[width=4.2cm, height=4.2cm]{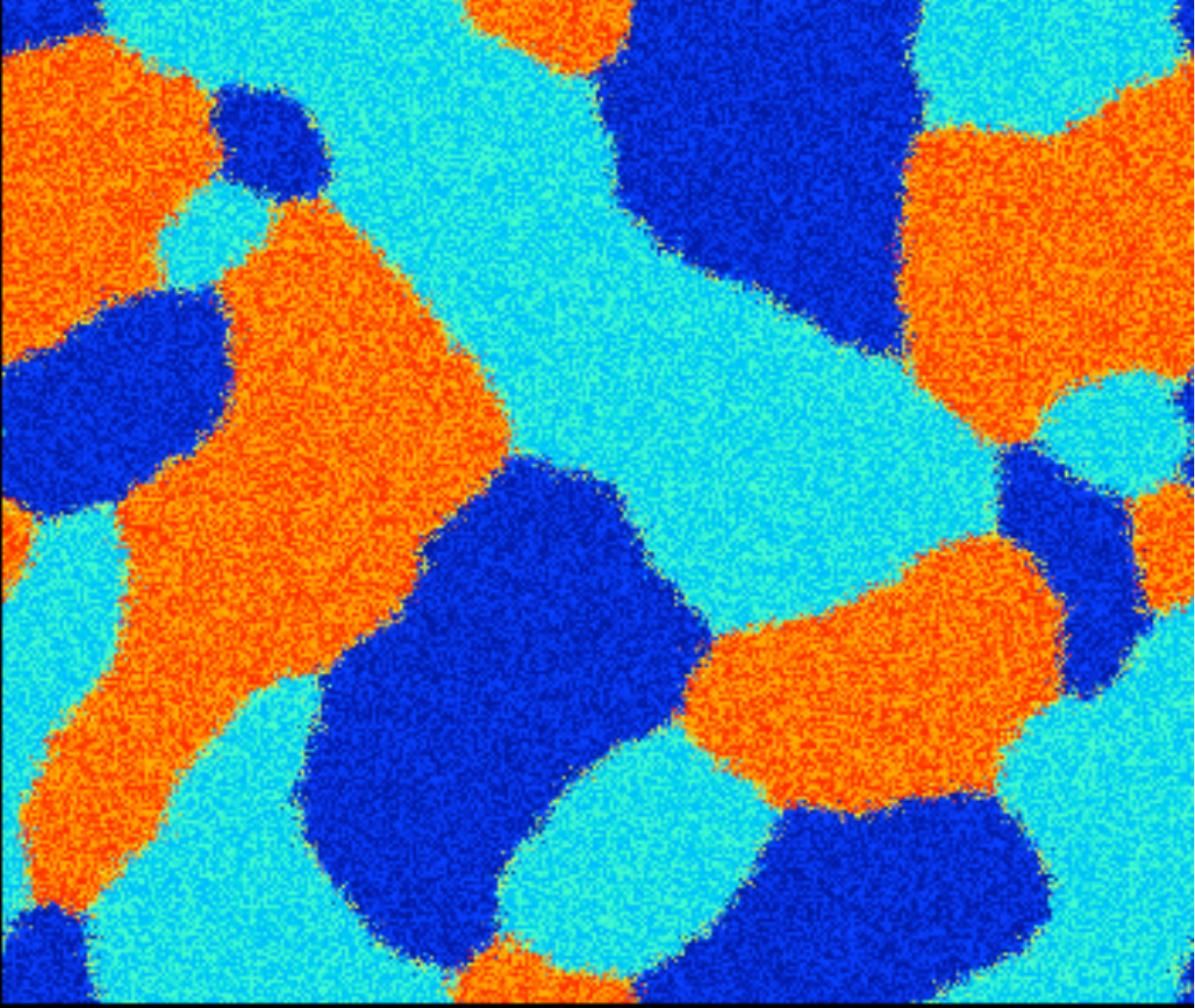}
\includegraphics*[width=4.2cm, height=4.2cm]{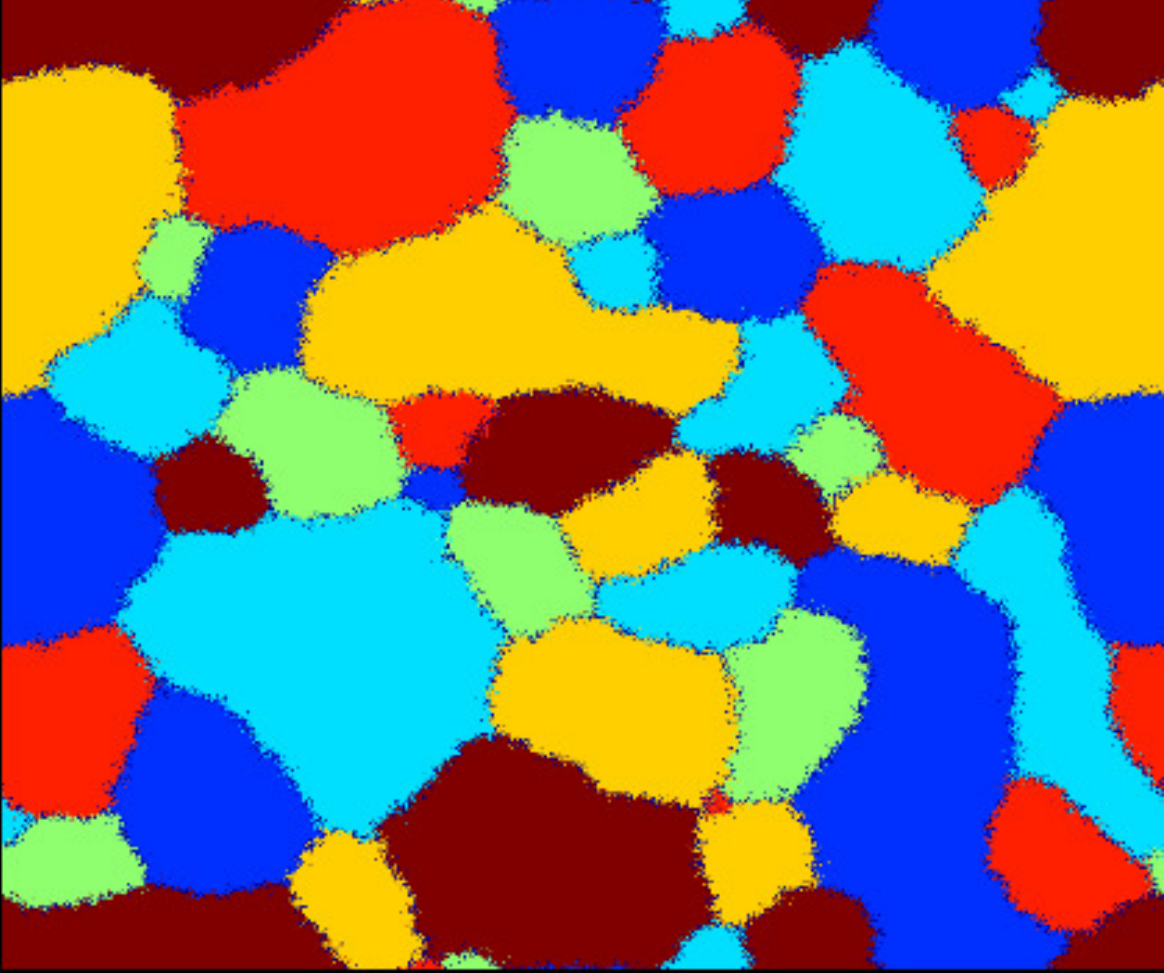}
\includegraphics*[width=4.2cm, height=4.2cm]{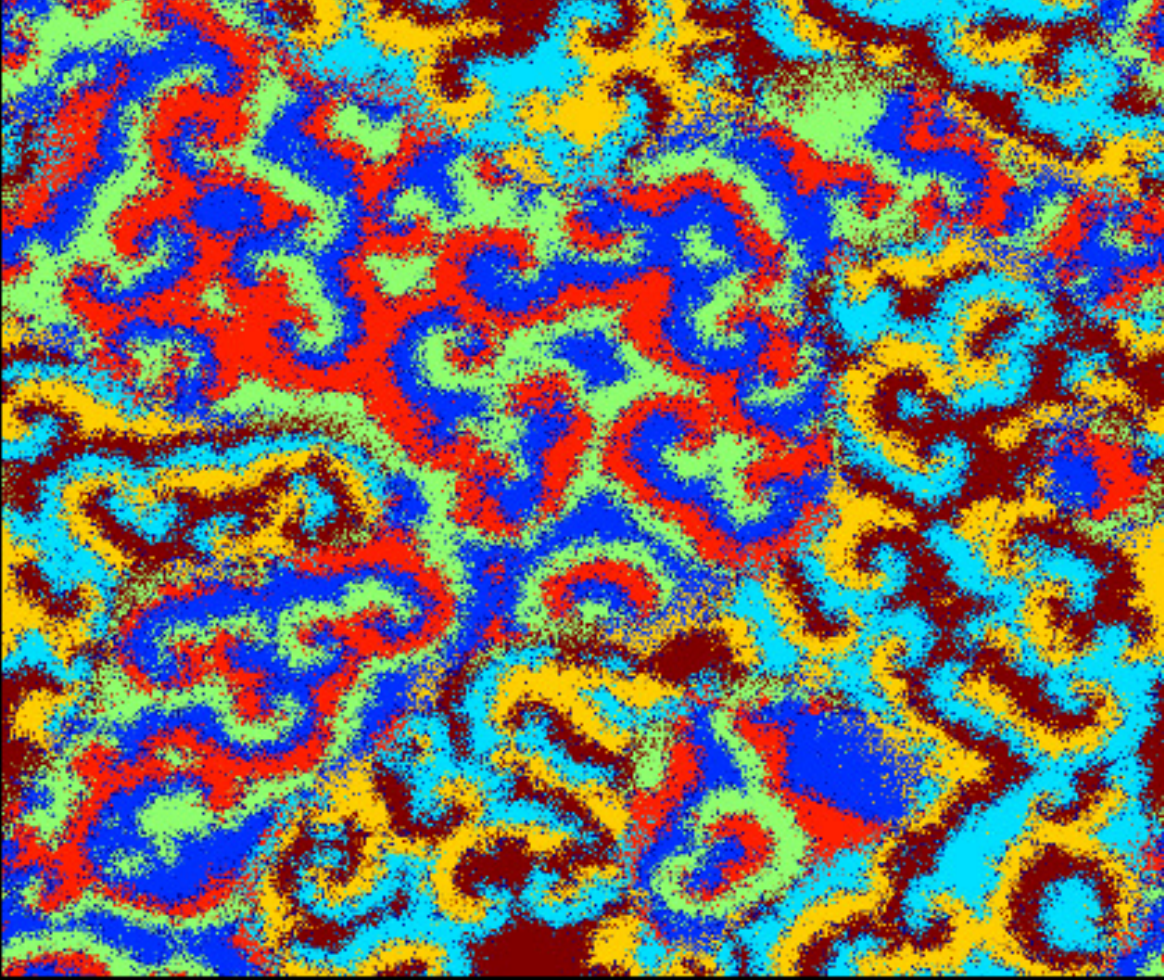}
\caption{(Color online) Snapshots of 4 different $600^2$ simulations with $N=6$ after $4000$ generations. The different panels show the results obtained using model II (upper left), model III (upper right), model IV (lower left) and model V (lower right).}
\end{figure}
%%%%%%%%%%%%%%%%%%%%%%%%%%%%%%%%%%%%%%%%%%%%%

Now we shall demonstrate that the curvature driven interface dynamics of model I is common in RPS type models. Let us consider a model with $N=6$. By taking $p_{L1}=p$ and  $p_{R1} = p_{L2} = p_{R2} = p_{L3} = p_{R3} = 0$ (model II) one ensures that the domains which appear in the simulations are populated with one of the following non-interacting species partnerships: $\{1, 3, 5\}$ and $\{2, 4, 6\}$. The interactions (of equal strength) between the 2 different partnership domains occur mainly at the border where a large number of empty spaces are continuously being generated (see video \cite{video1} and Fig. 6 (upper left)). If one considers the case with $p_{L1}=p_{L2} = p$,  $p_{R1}=p_{R2} = p_{L3}=p_{R3}=0$ (model III) then there are 3 possible partnership domains containing non-interacting species:  $\{1, 4\}$, $\{2, 5\}$ and $\{3, 6\}$ thus leading to an interface network with Y-type junctions (see video \cite{video2} and Fig. 6 (upper right)). In the case with $p_{L \alpha}=p_{R \alpha} = p$ for  $\alpha=1,2,3$ (model IV), there are no partnerships since every species is linked in a bidirectional way to all other species thus generating an interface between any 2 given domains. This gives rise to an interface network with Y-type junctions and 6 different domain types (see video \cite{video3} and Fig. 6 (lower left)). If one now takes $p_{L1}=p_{R1}=p_{L2} = p$ and $p_{R2}=p_{L3}=p_{R3}=0$ (model V) then there are 2 possible domains defined by $\{1, 3, 5\}$ and $\{2, 4, 6\}$ but due to the non-zero unidirectional probability $p_{R2}$ of interaction between species in the same domain, spiral patterns do form (see video \cite{video4} and Fig. 6 (lower right)). Inside the 2 different domains the interactions are those of the standard RPS model with 3 species. In the boundary there is an (average) equilibrium between the predation probabilities from individuals on either side of the wall (for example, individuals of the species 1 predate and are predated by individuals of the species 2 and 6). In models IV and V the colors light blue, dark blue, red, maroon, green and yellow represent species 1 to 6, respectively. This study represents a significant extension with respect to previous work and, to the best of our knowledge, it is the first time that models III, IV and V have been studied in the literature. Note that model V is very different from the  6 species extension of the 5 species Rock-Paper-Scissors-Lizard-Spock model proposed in \cite{Hawick_2011}, defined by $p_{R1}=p_{L2} = p$ and $p_{L1}=p_{R2}=p_{R3}=p_{L3}=0$ (see video \cite{video5}), whose dynamics is not curvature driven.

The scaling parameters $\lambda$ calculated numerically for all models are: $\lambda=-0.450 \pm 0.027$ (model I), $\lambda= -0.467 \pm 0.034$ (model II), $\lambda= -0.465 \pm 0.028$ (model III), $\lambda= -0.421 \pm 0.018$ (model IV) and $\lambda=-0.429 \pm 0.029$ (model V). In the case of model V, the empty spaces also appear associated to the spiral patterns. Hence, the parameter $\lambda$ for this model was obtained using only the empty spaces which have as some of the four immediate neighbors individuals from the 2 groups: $\{1, 3, 5\}$ and $\{2, 4, 6\}$. In all cases the scaling parameter $\lambda$ is reasonably close to the expected value $\lambda=-0.5$. The deviations can be attributed to the finite size and dynamical range of the simulations. We have verified that the results obtained for models I, II, III, and IV are weakly dependent on the value of $m$, as long as the thickness of the interfaces is much smaller than the box size. In the case of model V we have found that the network evolution is curvature dominated only if $m \gsim 0.5$. For smaller values of $m$ there are other effects which have a significant impact on the network dynamics, such as local partnerships along the borders of the domains \cite{PhysRevE.76.051921} (for example, between species 1 and 4 which do not predate each other). These effects are outside the scope of the present paper and will studied in detail in future work. In summary, we have shown that curvature driven interface dynamics, analogous to that observed in other physical systems in condensed matter and cosmology, is common in the family of RPS type models investigated in this paper and may be crucial to the understanding of biological complexity.

We thank Breno Oliveira for computational support and useful discussions. This work is partially supported by FCT-Portugal, CAPES and CNPq-Brazil.

\bibliography{ablm01}

\begin{thebibliography}{31}
\expandafter\ifx\csname natexlab\endcsname\relax\def\natexlab#1{#1}\fi
\expandafter\ifx\csname bibnamefont\endcsname\relax
  \def\bibnamefont#1{#1}\fi
\expandafter\ifx\csname bibfnamefont\endcsname\relax
  \def\bibfnamefont#1{#1}\fi
\expandafter\ifx\csname citenamefont\endcsname\relax
  \def\citenamefont#1{#1}\fi
\expandafter\ifx\csname url\endcsname\relax
  \def\url#1{\texttt{#1}}\fi
\expandafter\ifx\csname urlprefix\endcsname\relax\def\urlprefix{URL }\fi
\providecommand{\bibinfo}[2]{#2}
\providecommand{\eprint}[2][]{\url{#2}}

\bibitem[{\citenamefont{Kerr et~al.}(2002)\citenamefont{Kerr, Riley, Feldman,
  and Bohannan}}]{Kerr2002}
\bibinfo{author}{\bibfnamefont{B.}~\bibnamefont{Kerr}},
  \bibinfo{author}{\bibfnamefont{M.~A.} \bibnamefont{Riley}},
  \bibinfo{author}{\bibfnamefont{M.~W.} \bibnamefont{Feldman}},
  \bibnamefont{and} \bibinfo{author}{\bibfnamefont{B.~J.~M.}
  \bibnamefont{Bohannan}}, \bibinfo{journal}{Nature}
  \textbf{\bibinfo{volume}{418}}, \bibinfo{pages}{171} (\bibinfo{year}{2002}).

\bibitem[{\citenamefont{Reichenbach et~al.}(2007)\citenamefont{Reichenbach,
  Mobilia, and Frey}}]{Reichenbach2007}
\bibinfo{author}{\bibfnamefont{T.}~\bibnamefont{Reichenbach}},
  \bibinfo{author}{\bibfnamefont{M.}~\bibnamefont{Mobilia}}, \bibnamefont{and}
  \bibinfo{author}{\bibfnamefont{E.}~\bibnamefont{Frey}},
  \bibinfo{journal}{Nature} \textbf{\bibinfo{volume}{448}},
  \bibinfo{pages}{1046} (\bibinfo{year}{2007}).

\bibitem[{\citenamefont{Shi et~al.}(2010)\citenamefont{Shi, Wang, Yang, and
  Lai}}]{Shi2010}
\bibinfo{author}{\bibfnamefont{H.}~\bibnamefont{Shi}},
  \bibinfo{author}{\bibfnamefont{W.-X.} \bibnamefont{Wang}},
  \bibinfo{author}{\bibfnamefont{R.}~\bibnamefont{Yang}}, \bibnamefont{and}
  \bibinfo{author}{\bibfnamefont{Y.-C.} \bibnamefont{Lai}},
  \bibinfo{journal}{Phys. Rev. E} \textbf{\bibinfo{volume}{81}}
  (\bibinfo{year}{2010}).

\bibitem[{\citenamefont{Volterra}((1931))}]{Volterra}
\bibinfo{author}{\bibfnamefont{V.}~\bibnamefont{Volterra}},
  \emph{\bibinfo{title}{Lecons dur la Theorie Mathematique de la Lutte pour la
  Vie}} (\bibinfo{publisher}{Gauthier-Villars, Paris}, \bibinfo{year}{(1931)}),
  \bibinfo{edition}{$1^\mathrm{st}$} ed.

\bibitem[{\citenamefont{Lotka}(1920)}]{doi:10.1021/ja01453a010}
\bibinfo{author}{\bibfnamefont{A.~J.} \bibnamefont{Lotka}},
  \bibinfo{journal}{Journal of the American Chemical Society}
  \textbf{\bibinfo{volume}{42}}, \bibinfo{pages}{1595} (\bibinfo{year}{1920}).

\bibitem[{\citenamefont{Szab\'o et~al.}(2007)\citenamefont{Szab\'o, Szolnoki,
  and Sznaider}}]{PhysRevE.76.051921}
\bibinfo{author}{\bibfnamefont{G.}~\bibnamefont{Szab\'o}},
  \bibinfo{author}{\bibfnamefont{A.}~\bibnamefont{Szolnoki}}, \bibnamefont{and}
  \bibinfo{author}{\bibfnamefont{G.~A.} \bibnamefont{Sznaider}},
  \bibinfo{journal}{Phys. Rev. E} \textbf{\bibinfo{volume}{76}},
  \bibinfo{pages}{051921} (\bibinfo{year}{2007}).

\bibitem[{\citenamefont{Peltomaki and Alava}(2008)}]{Peltomaki2008}
\bibinfo{author}{\bibfnamefont{M.}~\bibnamefont{Peltomaki}} \bibnamefont{and}
  \bibinfo{author}{\bibfnamefont{M.}~\bibnamefont{Alava}},
  \bibinfo{journal}{Phys. Rev. E} \textbf{\bibinfo{volume}{78}}
  (\bibinfo{year}{2008}).

\bibitem[{\citenamefont{Szabo et~al.}(2008)\citenamefont{Szabo, Szolnoki, and
  Borsos}}]{Szabo2008}
\bibinfo{author}{\bibfnamefont{G.}~\bibnamefont{Szabo}},
  \bibinfo{author}{\bibfnamefont{A.}~\bibnamefont{Szolnoki}}, \bibnamefont{and}
  \bibinfo{author}{\bibfnamefont{I.}~\bibnamefont{Borsos}},
  \bibinfo{journal}{Phys. Rev. E} \textbf{\bibinfo{volume}{77}}
  (\bibinfo{year}{2008}).

\bibitem[{\citenamefont{Wang et~al.}(2010)\citenamefont{Wang, Lai, and
  Grebogi}}]{Wang2010}
\bibinfo{author}{\bibfnamefont{W.-X.} \bibnamefont{Wang}},
  \bibinfo{author}{\bibfnamefont{Y.-C.} \bibnamefont{Lai}}, \bibnamefont{and}
  \bibinfo{author}{\bibfnamefont{C.}~\bibnamefont{Grebogi}},
  \bibinfo{journal}{Phys. Rev. E} \textbf{\bibinfo{volume}{81}}
  (\bibinfo{year}{2010}).

\bibitem[{\citenamefont{Wang et~al.}(2011)\citenamefont{Wang, Ni, and
  Lai}}]{Wang2011}
\bibinfo{author}{\bibfnamefont{W.-X.} \bibnamefont{Wang}},
  \bibinfo{author}{\bibfnamefont{X.}~\bibnamefont{Ni}}, \bibnamefont{and}
  \bibinfo{author}{\bibfnamefont{C.}~\bibnamefont{Lai}, \bibfnamefont{Y.-C.
  an~Grebogi}}, \bibinfo{journal}{Phys. Rev. E} \textbf{\bibinfo{volume}{83}}
  (\bibinfo{year}{2011}).

\bibitem[{\citenamefont{Hawick}(2011{\natexlab{a}})}]{Hawick2011}
\bibinfo{author}{\bibfnamefont{K.}~\bibnamefont{Hawick}},
  \bibinfo{journal}{Proceedings of the IASTED International Conference on
  Modelling and Simulation} pp. \bibinfo{pages}{129--136}
  (\bibinfo{year}{2011}{\natexlab{a}}).

\bibitem[{\citenamefont{Hawick}(2011{\natexlab{b}})}]{Hawick_2011}
\bibinfo{author}{\bibfnamefont{K.~A.} \bibnamefont{Hawick}},
  \bibinfo{journal}{CSTN Computational Science Technical Note Series}
  (\bibinfo{year}{2011}{\natexlab{b}}),
  \urlprefix\url{http://www.massey.ac.nz/~kahawick/cstn/129/cstn-129.pdf}.

\bibitem[{\citenamefont{Zia}(2010)}]{1101.0018}
\bibinfo{author}{\bibfnamefont{R.~K.~P.} \bibnamefont{Zia}},
  \emph{\bibinfo{title}{General properties of a system of $s$ species competing
  pairwise}} (\bibinfo{year}{2010}), \eprint{arXiv:1101.0018}.

\bibitem[{\citenamefont{Dobrinevski and Frey}(2012)}]{PhysRevE.85.051903}
\bibinfo{author}{\bibfnamefont{A.}~\bibnamefont{Dobrinevski}} \bibnamefont{and}
  \bibinfo{author}{\bibfnamefont{E.}~\bibnamefont{Frey}},
  \bibinfo{journal}{Phys. Rev. E} \textbf{\bibinfo{volume}{85}},
  \bibinfo{pages}{051903} (\bibinfo{year}{2012}).

\bibitem[{\citenamefont{von Neumann}(1952)}]{vonNeumann1952}
\bibinfo{author}{\bibfnamefont{J.}~\bibnamefont{von Neumann}}, in
  \emph{\bibinfo{booktitle}{Metal Interfaces}} (\bibinfo{year}{1952}), pp.
  \bibinfo{pages}{108--110}.

\bibitem[{\citenamefont{Glazier}(1989)}]{Glazier-1989-PhdThesis}
\bibinfo{author}{\bibfnamefont{J.~A.} \bibnamefont{Glazier}},
  \bibinfo{type}{Phd thesis}, \bibinfo{school}{University of Chicago}
  (\bibinfo{year}{1989}),
  \urlprefix\url{http://biocomplexity.indiana.edu/jglazier/cv.php?pg=2}.

\bibitem[{\citenamefont{Stavans and Glazier}(1989)}]{Stavans1989}
\bibinfo{author}{\bibfnamefont{J.}~\bibnamefont{Stavans}} \bibnamefont{and}
  \bibinfo{author}{\bibfnamefont{J.~A.} \bibnamefont{Glazier}},
  \bibinfo{journal}{Phys. Rev. Lett.} \textbf{\bibinfo{volume}{62}},
  \bibinfo{pages}{1318} (\bibinfo{year}{1989}).

\bibitem[{\citenamefont{{Glazier} and {Weaire}}(1992)}]{Glazier1992}
\bibinfo{author}{\bibfnamefont{J.~A.} \bibnamefont{{Glazier}}}
  \bibnamefont{and} \bibinfo{author}{\bibfnamefont{D.}~\bibnamefont{{Weaire}}},
  \bibinfo{journal}{J. Phys. Cond. Mat.} \textbf{\bibinfo{volume}{4}},
  \bibinfo{pages}{1867} (\bibinfo{year}{1992}).

\bibitem[{\citenamefont{Flyvbjerg}(1993)}]{Flyvbjerg1993}
\bibinfo{author}{\bibfnamefont{H.}~\bibnamefont{Flyvbjerg}},
  \bibinfo{journal}{Phys. Rev. E} \textbf{\bibinfo{volume}{47}},
  \bibinfo{pages}{4037} (\bibinfo{year}{1993}).

\bibitem[{\citenamefont{{Monnereau} and {Vignes-Adler}}(1998)}]{Monnereau1998}
\bibinfo{author}{\bibfnamefont{C.}~\bibnamefont{{Monnereau}}} \bibnamefont{and}
  \bibinfo{author}{\bibfnamefont{M.}~\bibnamefont{{Vignes-Adler}}},
  \bibinfo{journal}{Phys. Rev. Lett.} \textbf{\bibinfo{volume}{80}},
  \bibinfo{pages}{5228} (\bibinfo{year}{1998}).

\bibitem[{\citenamefont{Weaire and Hutzler}(2000)}]{Weaire2000}
\bibinfo{author}{\bibfnamefont{D.}~\bibnamefont{Weaire}} \bibnamefont{and}
  \bibinfo{author}{\bibfnamefont{R.}~\bibnamefont{Hutzler}},
  \emph{\bibinfo{title}{The physics of foams}} (\bibinfo{publisher}{Oxford
  University Press}, \bibinfo{address}{Oxford}, \bibinfo{year}{2000}).

\bibitem[{\citenamefont{{Kim} et~al.}(2006)\citenamefont{{Kim}, {Kim}, {Kim},
  and {Park}}}]{Kim2006}
\bibinfo{author}{\bibfnamefont{S.~G.} \bibnamefont{{Kim}}},
  \bibinfo{author}{\bibfnamefont{D.~I.} \bibnamefont{{Kim}}},
  \bibinfo{author}{\bibfnamefont{W.~T.} \bibnamefont{{Kim}}}, \bibnamefont{and}
  \bibinfo{author}{\bibfnamefont{Y.~B.} \bibnamefont{{Park}}},
  \bibinfo{journal}{\pre} \textbf{\bibinfo{volume}{74}},
  \bibinfo{pages}{061605} (\bibinfo{year}{2006}).

\bibitem[{\citenamefont{Arenzon et~al.}(2007)\citenamefont{Arenzon, Bray,
  Cugliandolo, and Sicilia}}]{PhysRevLett.98.145701}
\bibinfo{author}{\bibfnamefont{J.~J.} \bibnamefont{Arenzon}},
  \bibinfo{author}{\bibfnamefont{A.~J.} \bibnamefont{Bray}},
  \bibinfo{author}{\bibfnamefont{L.~F.} \bibnamefont{Cugliandolo}},
  \bibnamefont{and} \bibinfo{author}{\bibfnamefont{A.}~\bibnamefont{Sicilia}},
  \bibinfo{journal}{Phys. Rev. Lett.} \textbf{\bibinfo{volume}{98}},
  \bibinfo{pages}{145701} (\bibinfo{year}{2007}).

\bibitem[{\citenamefont{Avelino et~al.}(2008)\citenamefont{Avelino, Martins,
  Menezes, Menezes, and Oliveira}}]{Avelino2008}
\bibinfo{author}{\bibfnamefont{P.~P.} \bibnamefont{Avelino}},
  \bibinfo{author}{\bibfnamefont{C.~J. A.~P.} \bibnamefont{Martins}},
  \bibinfo{author}{\bibfnamefont{J.}~\bibnamefont{Menezes}},
  \bibinfo{author}{\bibfnamefont{R.}~\bibnamefont{Menezes}}, \bibnamefont{and}
  \bibinfo{author}{\bibfnamefont{J.~C. R.~E.} \bibnamefont{Oliveira}},
  \bibinfo{journal}{Phys. Rev. D} \textbf{\bibinfo{volume}{78}},
  \bibinfo{pages}{103508} (\bibinfo{year}{2008}).

\bibitem[{\citenamefont{Avelino et~al.}(2011)\citenamefont{Avelino, Menezes,
  and Oliveira}}]{Avelino2010}
\bibinfo{author}{\bibfnamefont{P.~P.} \bibnamefont{Avelino}},
  \bibinfo{author}{\bibfnamefont{R.}~\bibnamefont{Menezes}}, \bibnamefont{and}
  \bibinfo{author}{\bibfnamefont{J.~C. R.~E.} \bibnamefont{Oliveira}},
  \bibinfo{journal}{Phys. Rev. E} \textbf{\bibinfo{volume}{83}},
  \bibinfo{pages}{011602} (\bibinfo{year}{2011}).

\bibitem[{\citenamefont{Avelino et~al.}(2005)\citenamefont{Avelino, Oliveira,
  and Martins}}]{Avelino2005}
\bibinfo{author}{\bibfnamefont{P.~P.} \bibnamefont{Avelino}},
  \bibinfo{author}{\bibfnamefont{J.~C. R.~E.} \bibnamefont{Oliveira}},
  \bibnamefont{and} \bibinfo{author}{\bibfnamefont{C.~J. A.~P.}
  \bibnamefont{Martins}}, \bibinfo{journal}{Phys. Lett. B}
  \textbf{\bibinfo{volume}{610}}, \bibinfo{pages}{1} (\bibinfo{year}{2005}).

\bibitem[{vid({\natexlab{a}})}]{video1}
\urlprefix\url{http://www.youtube.com/watch?v=Q3Y43kKHC_8}.

\bibitem[{vid({\natexlab{b}})}]{video2}
\urlprefix\url{http://www.youtube.com/watch?v=LeAz2i2FUu8}.

\bibitem[{vid({\natexlab{c}})}]{video3}
\urlprefix\url{http://www.youtube.com/watch?v=kZRWQwDNq70}.

\bibitem[{vid({\natexlab{d}})}]{video4}
\urlprefix\url{http://www.youtube.com/watch?v=__jSZOyMtxE}.

\bibitem[{vid({\natexlab{e}})}]{video5}
\urlprefix\url{http://www.youtube.com/watch?v=jFEmm7W3b2Q}.

\end{thebibliography}

\end{document}